\documentclass{jpsj-suppl}
\usepackage{txfonts} 
\usepackage{amsmath,amssymb}
\usepackage{bm}
\usepackage{color}
\usepackage{longtable}

\title{Construction of $\bar{K}N$ potential and structure of $\Lambda(1405)$ \\ based on chiral unitary approach}

\author{Kenta \textsc{Miyahara}$^{1}$ and Tetsuo \textsc{Hyodo}$^{2}$}

\inst{$^{1}$Department of Physics, Graduate School of Science, Kyoto University, Kyoto 606-8502, Japan \\
$^{2}$Yukawa Institute for Theoretical Physics, Kyoto University, Kyoto 606-8502, Japan}

\email{miyahara@ruby.scphys.kyoto-u.ac.jp}

\recdate{August 23, 2015}

\abst{Based on chiral unitary approach, we construct the realistic $\bar{K}N$ local potential, which is useful for the quantitative calculation of $\bar{K}$-nuclei. Since the resonance pole structure of the $\bar{K}N$ system seems important for the $\bar{K}$-nuclei and the spacial structure of $\Lambda(1405)$, we establish the construction procedure of the local potential paying attention to the scattering amplitude in the complex energy plane. Furthermore, for the quantitative study of the $\bar{K}$-nuclei, we consider the constraint from the recent experimental data measured by SIDDHARTA, which significantly reduces the uncertainty of the $\bar{K}N$ amplitude. With this new local potential, we estimate the spacial structure of $\Lambda(1405)$ and obtain the result indicating the meson-baryon molecular state of $\Lambda(1405)$.}

\kword{$\bar{K}N$ local potential, $\Lambda(1405)$, chiral unitary approach, complex energy plane, SIDDHARTA data}

\begin{document}
\maketitle

\section{Introduction}
Recently, the study of the multi-nucleon systems with an antikaon have been active, since the strong attraction in the $\bar{K}N$ channel is considered to lead to various interesting phenomena. The simplest example is the $\Lambda(1405)$ resonance \cite{Dalitz:1959dn,Dalitz:1960du}. 
Since it is difficult to describe $\Lambda(1405)$ by the conventional three-quark picture in the constituent quark model, this is considered to be the quasi-bound state of the $\bar{K}N$ system slightly below the threshold.
Another example is the $\bar{K}$-nuclei \cite{Akaishi:2002bg,Yamazaki:2002uh}, which may have the  exotic structure because of the strong $\bar{K}N$ attraction. 
Quantitatively, the predictions of the mass and the width of the $\bar{K}N$-nuclei are substantially different from each other. 
This is because the $\bar{K}N$ interaction below the $\bar{K}N$ threshold, which is essential for the calculations of the $\bar{K}$-nuclei, has not been constrained strongly from previous experimental data. 
Recently, the precise $K^-p$ scattering amplitude has been obtained from the SIDDHARTA data \cite{Bazzi:2011zj,Bazzi:2012eq}, which reduces the subthreshold uncertainty of the $\bar{K}N$ interaction significantly. 

For a quantitative prediction of $\Lambda(1405)$ and the $\bar{K}$-nuclei with such precise data, we have to establish the reliable construction procedure of the hadron interaction. 
In this work, we construct the $\bar{K}N$ local potential which is useful for the few-body calculations \cite{Miyahara:2015bya}. The potential is constructed to reproduce the scattering amplitude from chiral unitary approach, which is based on chiral perturbation theory and unitarity of the scattering amplitude, following Ref.~\cite{Hyodo:2007jq}. 
To analyze the structure of $\Lambda(1405)$, it is necessary to reproduce the amplitude in the complex energy plane. Hence, paying attention to the precision in the complex plane and using the precise SIDDHARTA data, we construct the reliable $\bar{K}N$ potential applicable for the quantitative calculations. Furthermore, with this new $\bar{K}N$ potential, we investigate the spacial structure of $\Lambda(1405)$.

\section{Formulation}
To describe the $\bar{K}N$ scattering, we utilize the chiral unitary approach \cite{Kaiser:1995eg,Oset:1997it,Oller:2000fj,Lutz:2001yb,Hyodo:2011ur}. In the $s$-wave meson-baryon system, the scattering amplitude $T_{ij}(\sqrt{s})$ at the total center of mass energy $\sqrt{s}$ is
\begin{align}
T_{ij}(\sqrt{s}) &= V_{ij}(\sqrt{s}) + V_{ik}(\sqrt{s})G_{k}(\sqrt{s})T_{kj}(\sqrt{s})    \label{eq:BetheSal} \\
&= \left[ ({V(\sqrt{s})}^{-1}-G(\sqrt{s}))^{-1} \right]_{ij},  \notag
\end{align}
where $V_{ij}$ and $G_i$ are respectively the meson-baryon interaction kernel from chiral perturbation theory and the loop function with the meson-baryon channel indices denoted by $i,j$. 
The amplitude $T_{ij}$ is related to the $\bar{K}N$ forward scattering amplitude $F_{\bar{K}N}= -M_N/(4\pi\sqrt{s})T_{11}(\sqrt{s})$, where $M_N$ represents the nucleon mass. 
The amplitude has two resonance poles of $\Lambda(1405)$ in the isospin $I=0$ channel. These poles are induced by the attractive interactions of the $\bar{K}N$ and $\pi\Sigma$ channels \cite{Hyodo:2007jq,Jido:2003cb}. Here, we call the higher (lower) energy pole $\bar{K}N$ pole ($\pi\Sigma$ pole).

What we want here is the single-channel $\bar{K}N$ local potential, which is easy to apply to the few-body systems. We construct this local potential in such a way as to produce the amplitude equivalent to the chiral coupled-channel approach. 
To this end, we use the single-channel $\bar{K}N$ interaction $V^{\rm eff}_{11}$ extracted by the Feshbach projection. Because the lower energy $\pi\Sigma$ channel is eliminated, the effective $\bar{K}N$ interaction $V^{\rm eff}_{11}$ becomes a complex valued function.
We first construct the energy dependent local potential $U(r,E)$,
\begin{align}
U(r,E) &= g(r)\frac{M_N}{2(E+M_N+m_K)}\frac{\omega_K+E_N}{\omega_K E_N}V^{\rm eff}_{11}(E+M_N+m_K),  \label{eq:equivpot0}
\end{align}
where $E$, $E_N$ and $\omega_K$ represent the nonrelativistic energy, the energy of the nucleon and the energy of the anti-kaon. The spatial structure of the potential is decided by $g(r)$ which is chosen as a Gaussian, $g(r) = 1/(\pi^{3/2}b^3)\ e^{-r^2/b^2}$, where $b$ is the range parameter. The flux factor depending on the energy is determined to reproduce the original amplitude at the $\bar{K}N$ threshold in the Born approximation \cite{Hyodo:2007jq}. 
Using the local potential, we can calculate the wave function from the Schr\"odinger equation. 
From the behavior of the wave function at $r\to\infty$, the scattering amplitude $F_{\bar{K}N}$ can be obtained. 
In Ref.~\cite{Hyodo:2007jq}, the parameter $b$ was determined to reproduce the original amplitude in the $\Lambda(1405)$ resonance. In this work, we determine $b$ to reproduce the original amplitude at the $\bar{K}N$ threshold, which is consistent with the determination of the flux factor.

The previous potential~\eqref{eq:equivpot0} well reproduces the original amplitude on the real energy axis near the $\bar{K}N$ threshold. However, the deviation becomes large far below the threshold. To extend the reliable energy region of the potential, we add the correction $\Delta V(E)$ to the potential,
\begin{align}
U(r,E) &= g(r)\frac{M_N}{2(E+M_N+m_K)}\frac{\omega_K+E_N}{\omega_K E_N}\left[V^{\rm eff}_{11}(E+M_N+m_K)+\Delta V(E) \right].  \label{eq:equivpot1}
\end{align}
We parameterize the strength of the potential by a polynomial in the energy for the analytic continuation of the amplitude in the complex energy plane, and refer to the energy range where the potential is parameterized as parameterized range.

\section{Potential construction}

\begin{figure}[tb]
\begin{center}
\includegraphics[width=8cm,bb=240 0 750 311]{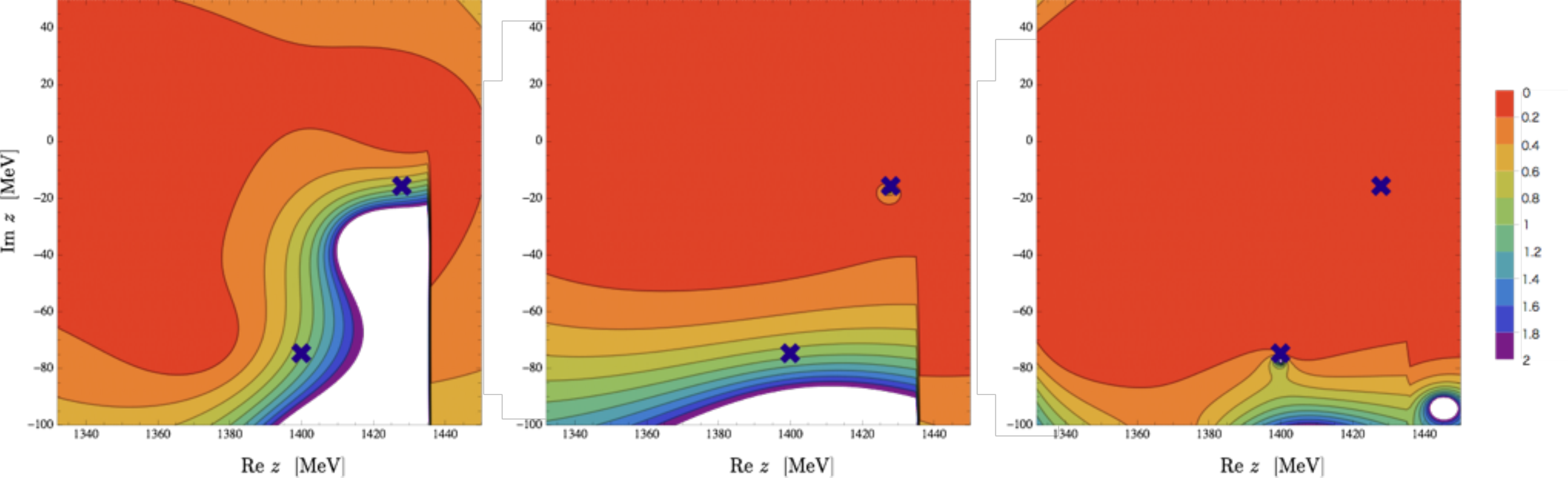}
\caption{The contour plot of $\Delta F(z)\equiv \left| \left(F_{\bar{K}N}^{\rm Ch}(z)-F_{\bar{K}N}(z)\right)/F_{\bar{K}N}^{\rm Ch}(z) \right|$ of the previous potential in Ref.~\cite{Hyodo:2007jq}, Potential I, and Potential II. The unfilled region corresponds to large deviation, $\Delta F>2$. Precise region is defined as $\Delta F<0.2$. The crosses represent the original pole positions of $\Lambda(1405)$. }
\label{fig:prec}  
\end{center}
\end{figure}

The previous potentials in Ref.~\cite{Hyodo:2007jq} well reproduce the $\bar{K}N$ amplitudes on the real axis. On the other hand, we find that the amplitudes from the local potentials $F_{\bar{K}N}$ in the complex energy plane are quite different from the original amplitudes directly from chiral unitary approaches $F^{\rm Ch}_{\bar{K}N}$ as shown in Fig.~\ref{fig:prec}. This figure shows the contour plot of the deviation between $F_{\bar{K}N}$ and $F^{\rm Ch}_{\bar{K}N}$, $\Delta F(z) \equiv \left| \left(F_{\bar{K}N}^{\rm Ch}(z)-F_{\bar{K}N}(z)\right)/F_{\bar{K}N}^{\rm Ch}(z) \right|$, where $z$ represents the complex energy of the two-body system. 
Related to the deviation in the complex plane, the resonance poles of $\Lambda(1405)$ are also different from the original ones (the pole position from the local potential is $1421-35i$ MeV. On the other hand, the original pole positions are $1428-17i$ MeV and $1400-76i$ MeV). Hence, we should establish the improved construction procedure of the local potential paying attention to the amplitude in the complex plane.

First, we focus on the energy region near the real energy axis around the $\bar{K}N$ pole.
From Fig.~\ref{fig:prec}, we find that the deviation around the $\Lambda(1405)$ energy region is larger than the other region. This is because in the previous work, $\Delta V$ in Eq.~\eqref{eq:equivpot1} was added only in the region below $1400\ {\rm MeV}$. Furthermore, $\Delta V$ was a real number, considering that the real part in $V^{\rm eff}_{11}$ is larger than the imaginary part.
In this work, we add $\Delta V$ in the region around the $\bar{K}N$ pole, 1332-1450 MeV, and introduce the complex correction $\Delta V$ to reproduce the original amplitude near the $\Lambda(1405)$ resonance region. 
For the estimation of the improvement, we define the average deviation $\Delta F_{\rm real}$ between the amplitude from chiral unitary approach and from the local potential on the real axis,
\begin{align}
\Delta F_{\rm real} = \frac{\displaystyle\int d\sqrt{s}\ |F_{\bar{K}N}^{\rm Ch}(\sqrt{s})-F_{\bar{K}N}(\sqrt{s})|}{\displaystyle\int d\sqrt{s}\ |F_{\bar{K}N}^{\rm Ch}(\sqrt{s})|}. \notag
\end{align}
As a result of the above improvements, $\Delta F_{\rm real}$ is largely reduced from $1.4\times10^{-1}$ to $4.8\times10^{-3}$, which leads to the improvement of the $\bar{K}N$ pole position (the position is $1421-35i$ MeV from the previous potential, and $1427-17i$ MeV after the improvements, in comparison with the original position,$1428-17i$ MeV). We call the new potential ``Potential I'' and the contour plot of $\Delta F$ in the complex plane is shown in Fig.~\ref{fig:prec}. From this figure, we find that the ``precise region" satisfying $\Delta F<0.2$ is extended over the $\bar{K}N$ pole.

Though the $\bar{K}N$ pole is reproduced by the above improvements, the $\pi\Sigma$ pole far from the real axis does not appear. Therefore, we add another improvement to reproduce the amplitude far from the real axis.
In principle, if we completely reproduce the original amplitude in the whole range on the real energy axis, the amplitude in the complex plane should be same as the original one. Hence, we broaden the parameterized range and increase the degree of the polynomial from the third order to the tenth order.
We test the various parametrized ranges and find the best potential which almost reproduces the original $\bar{K}N$ and $\pi\Sigma$ poles, $1428-17i$ MeV and $1400-76i$ MeV (the pole positions from the local potential are $1428-17i$ MeV and $1400-77i$ MeV). We call this potential ``Potential II" and the contour plot of $\Delta F$ in the complex plane is shown in Fig.~\ref{fig:prec}. From this figure, we find that the precise region is extended near the $\pi\Sigma$ pole. 
\section{Application}

\begin{figure}[tb]
\begin{center}
\includegraphics[width=8cm,bb=0 0 603 489]{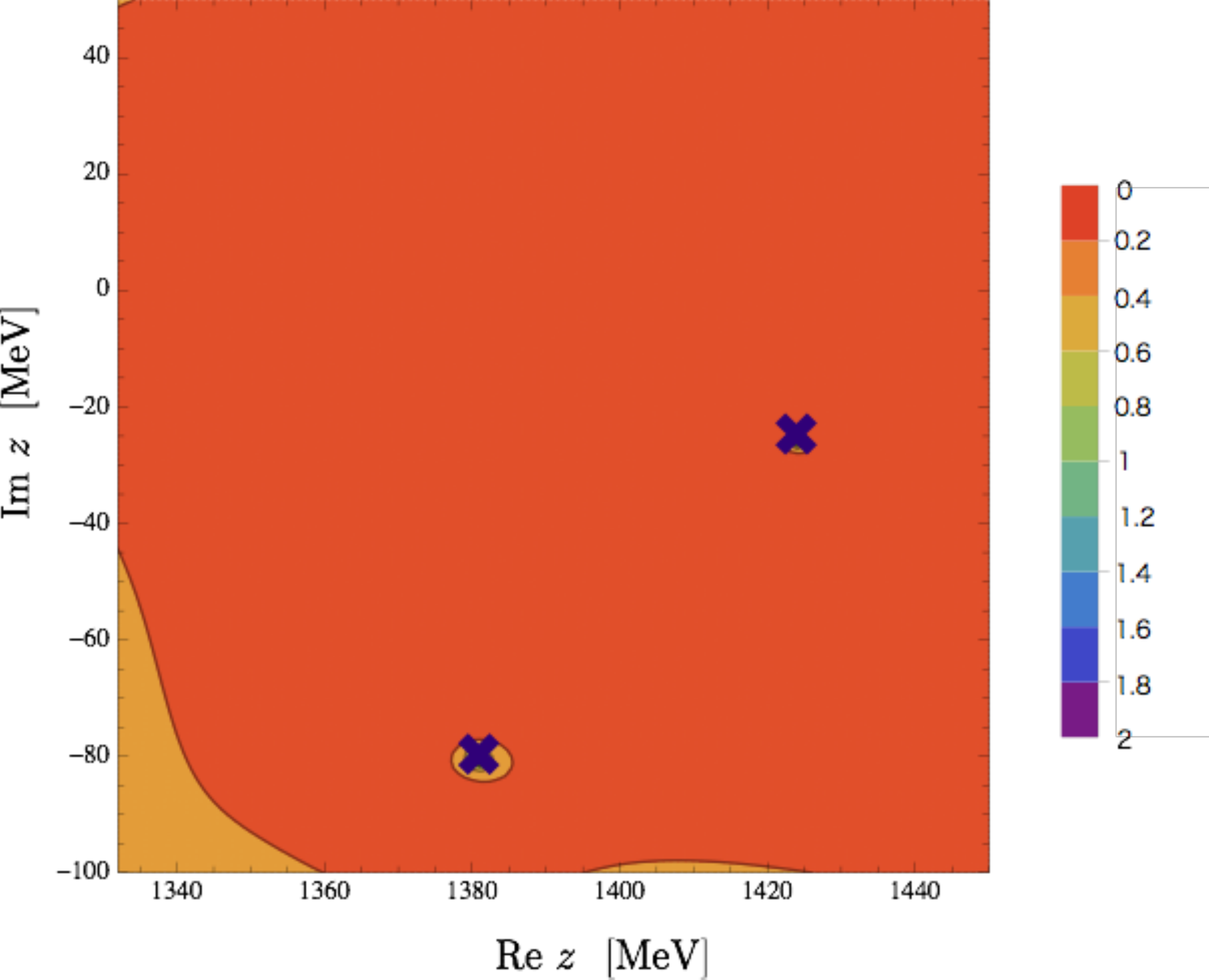}
\caption{The contour plot of $\Delta F$ of SIDDHARTA potential $(I=0)$. Precise region is defined as $\Delta F<0.2$. The crosses represent the original pole positions of $\Lambda(1405)$. }
\label{fig:sidprec}  
\end{center}
\end{figure}

Applying the above construction procedure of the local potential to the amplitude of Refs.~\cite{Ikeda:2011pi,Ikeda:2012au}, we constructed the most reliable $\bar{K}N$ potential with the SIDDHARTA data. We call the potential in isospin $I=0$ SIDDHARTA potential $(I=0)$. The contour plot of $\Delta F$ is shown in Fig.~\ref{fig:sidprec}. From this figure, it is seen that SIDDHARTA potential $(I=0)$ well reproduces the original amplitude in the complex energy plane. As a result, we obtain the poles of $\Lambda(1405)$ at the same position as the original ones in the accuracy of 1 MeV.
%

In the same way, the $\bar{K}N$ local potential for the $I=1$ channel, SIDDHARTA potential $(I=1)$, is constructed. With SIDDHARTA potential $(I=0)$ and SIDDHARTA potential $(I=1)$, we can calculate few-body systems. 

Finally, using SIDDHARTA potential $(I=0)$, we estimate the $\bar{K}N$ distance, that is, the spacial structure of $\Lambda(1405)$. Solving Schr\"odinger equation, we obtain the $\bar{K}N$ wave function in $s$ wave. With this wave function, we calculate the root mean squared distance, and obtain the result, 1.44 fm. Comparing with the charged radii of the proton (0.85 fm) and $K^-$ (0.55 fm), the $\bar{K}N$ distance is relatively larger, which means the meson-baryon molecular state of $\Lambda(1405)$.



\end{document}